\documentclass{amsart}

\usepackage{pstricks,color,amsthm, epsfig, amssymb}

\newtheorem{thm}{Theorem}

\newtheorem{lemma}[thm]{Lemma}
\newtheorem{prop}[thm]{Proposition}

\newcommand{\be}{\begin{equation}}
\newcommand{\ee}{\end{equation}}
\newcommand{\bea}{\begin{eqnarray}}
\newcommand{\eea}{\end{eqnarray}}
\newcommand{\beas}{\begin{eqnarray*}}
\newcommand{\eeas}{\end{eqnarray*}}

\def\One{\mathbb{I}}

\def\paa{\;\raisebox{-2mm}{\epsfysize=6mm\epsfbox{p11.epsi}}\;}
\def\pba{\;\raisebox{-2mm}{\epsfysize=6mm\epsfbox{p21.epsi}}\;}
\def\pbb{\;\raisebox{-2mm}{\epsfysize=6mm\epsfbox{p22.epsi}}\;}
\def\pbc{\;\raisebox{-2mm}{\epsfysize=6mm\epsfbox{p23.epsi}}\;}

\title[Recursive estimates in QFT]{Recursion and growth estimates in renormalizable quantum field
  theory$^{*}$}

\author[Dirk Kreimer]{Dirk Kreimer$^{1}$}
\address{IH\'ES, Le Bois-Marie, 35, route de Chartres, F-91440
  Bures--sur--Yvette, FRANCE \\and Department of Mathematics and
  Statistics, Boston University,
 111 Cummington Street, Boston, MA 02215, USA}
\email{kreimer@ihes.fr}
\author{Karen Yeats}
\address{Department of Mathematics and Statistics, Boston University,
 111 Cummington Street, Boston, MA 02215, USA}
\email{kayeats@bu.edu}

\date{December 17, 2006.  BU--CMP/06--05}
\thanks{$^{*}$Research supported by grant NSF-DMS/0603781. $^{1}$supported by CNRS}
\begin{document}

\begin{abstract}
In this paper we show that there is a Lipatov bound for the radius of convergence
for superficially divergent one-particle irreducible Green functions in a renormalizable quantum field theory
if there is such a bound for the superficially convergent ones. The radius of convergence turns out to be ${\rm min}\{\rho,1/b_1\}$,
where $\rho$ is the bound on the convergent ones, the instanton radius, and $b_1$ the first coefficient of the $\beta$-function.
\end{abstract}
\maketitle
\section{The set-up}\subsection{Introduction}
In this paper we explore the recursive
structure of the short-distance sector of a renormalizable quantum
field theory. Such theories have a finite number of distinct amplitudes $r\in {\mathcal R}\subset {\mathcal A}$ which need renormalization.
We decompose each Green function in accordance with structure functions which need renormalization, the remaining structure functions and superficially convergent Green functions contribute to amplitudes $a\in {\mathcal A}\smallsetminus{\mathcal R}$ in the set of all amplitudes ${\mathcal A}$ \cite{dktor}.

For such a theory, the Dyson--Schwinger equations give
any amplitude $s\in {\mathcal A}={\mathcal A}\smallsetminus{\mathcal R}\cup{\mathcal R}$  in terms of all those amplitudes $r\in {\mathcal R}$
in the theory which need renormalization, and in terms of an
infinite series of integral kernels, the skeleton graphs of the theory. The very fact that we can
renormalize by modifying the Lagrangian implies that we have a sub
Hopf algebra at our disposal which has the graded elements $c_k^r$
as generators which correspond to the $k$-th order contribution to the
amplitude $r$ \cite{non lin,dktor}.

We can relate the integral kernels above to the primitives of the
Hopf algebra structure underlying renormalization. The growth of
these primitives is hence determined by the growth of integral
kernels provided by overall convergent Green functions. A bound on such a growth
is hence obtainable from a bound of amplitudes in ${\mathcal A}\smallsetminus{\mathcal R}$,
where results in constructive field theory are principally available.
The typical example is quantum electrodynamics (QED), where the primitives for the vertex function
are given by the superficially convergent four fermion $e^+e^-\to e^+ e^-$ scattering kernel, two-particle irreducible
 in a suitable channel.

To proceed from there to a bound for amplitudes in ${\mathcal R}$,
we need a handle on the behaviour of the singular integrations which
encompass the short-distance singular sector. Here, we proceed by
chosing a suitable set of primitives such that we can reduce the
Dyson--Schwinger equations to recursive equations acting on
one-variable Mellin transforms. A simple study of the conformal
symmetries in the corresponding primitives suffice to determine the
form of these Mellin transforms.

For any $r\in{\mathcal R}$, we can define a Green function \be
G^r(a,L)=1\pm\sum_k \gamma_k^r(a)L^k,\ee where $L=\ln -Q^2/\mu^2$
measures the scaling behaviour under the renormalization group
flow with respect to to a single Euclidean kinematical variable $Q^2<0$.

Hence, if we have a recursive set of equations giving Green
fuctions  $G^r(a,L)$ in terms of themselves, inserted into integral
kernels, we can apply Mellin transforms as defined below upon using \be G^r(a,\ln
-k^2/\mu^2)=G^r(a,\partial_\rho)\left(-\frac{k^2}{\mu^2}\right)^{\rho}\bigg|_{\rho=0},\ee
to reduce the evaluation of all integral kernels above to a study
of corresponding Mellin transforms.

In such a situation, we hence show how the growth of superficially
divergent amplitudes is related to the growth of the superficially
convergent ones. To the extent the latter is under control by
results in constructive field theory, we get results for the former.

The representation of primitive integral kernels through their
Mellin transform then allows us to turn the Dyson--Schwinger
equations into a recursive system which determines the $\gamma_k$ in
terms of the $\gamma_1$, and determines $\gamma_1(a)$ recursively
through the Taylor coefficients of the relevant Mellin transforms.
Modest knowledge of the structure of those transforms: \be
M(p)(\rho)=\frac{{\rm res}_p}{\rho(1-\rho)},\ee and a Lipatov bound
for them, $a^{|p|}{\rm res}_p\sim |p|!c^{|p|}$ for some suitable $c$
at large $|p|$, allows us to show that solutions to Dyson--Schwinger
equations for Green functions $G^r(a,L)$ have a similar Lipatov
bound.

We emphasize that our construction of a basis of primitives with a
given Mellin transform resolves overlapping divergences, thanks to
the Hochschild cohomology of the relevant Hopf algebras
\cite{OnOverl}.

We have thus a self-similar recursive
system determining the formal sums $\Gamma^r$, $r\in {\mathcal R}$
in terms of themselves and the action of suitable maps
$B_+^{k;r}=\sum_{|p|=k}B_+^{p;r}$ \cite{dktor}. For all $r\in{\mathcal R}$,
\be X^r(a)=\One\pm\sum_{k\geq 1}a^k B_+^{k;r}(X^r{\mathcal Q}^k),\ee
where $a{\mathcal Q}$ is the invariant charge of the theory, and Green functions are obtained as
\be G(a,L)=\phi_R(X^r(a))(L),\ee for renormalized Feynman rules $\phi_R$.

The study of the Hochschild one-cocycles $B_+^{k,i;r}$ is crucial
for a QFT. The Hopf algebra elements $B_+^{k;r}(\One)$ provide the
very integral kernels above underlying the DSEs and are the terms
which drive the recursion. Their consistent construction
automatically takes internal symmetries of the theory into
account, \cite{anatomy}, by dividing the Hopf algebra by suitable
Hopf ideals \cite{WvS1,WvS2,anatomy}. A further ideal is defined
by our desire to concentrate on amplitudes in ${\mathcal R}$.
\subsection{Ideals}
Green functions in field theory decompose into structure functions
which have logarithmic short-distance singularities. Locality of
field theory guarantees that these short distance singularities
are invariant under changes of dimensionfull parameters in the
theory, allowing for local counterterms. This is not the only
invariance which we can observe in counterterms. A chosen subgraph
needs the same counterterm wherever it appear in a larger graph,
and in whatever orientation it is inserted, as long as we work in
a symmetric renormalization scheme. Furthermore, its counterterm
is invariant under modification of its external momenta. We can
hence isolate short-distance singularities in subgraphs in a
manner such that all subdivergences are functions of a chosen
internal momentum of the cograph under consideration.

In so doing we enlarge the  Hopf algebra of graphs into a Hopf algebra of
colored graphs or decorated rooted trees, which makes the underlying Hochschild
cohomology obvious and resolves overlapping divergences
\cite{OnOverl,old}.

We emphasize that the choices above depend on the theory, renormalization scheme and
physical problem one wants to study. Here, we are only concerned with the general set-up, and
will only sporadically restrict to a specific theory. Our final result applies to any renormalizable field theory.
Specifics of any particular such theory are to be discussed elsewhere.

The freedom we have in the above choices can be summarized by saying
we work in an ideal  given by \be \sum_i
\Gamma^{\prime\prime}\circ_i {\rm End}^{-1}\Gamma^\prime=0, \ee with
\be {\rm End}(\Gamma_1\Gamma_2)={\rm End}(\Gamma_1){\rm
End}(\Gamma_2).\ee Here, we use Sweedler notation
$\Delta(\Gamma)=\sum_i\Gamma^{\prime}\otimes\Gamma^{\prime\prime}$
and the fact that Feynman graphs have an operadic structure, so that
for each term in the sum we have an operadic composition \cite{old}
of graphs such that \be
\Gamma=\Gamma^{\prime\prime}\circ_i\Gamma^{\prime}.\ee Finally,
${\rm End}$ is a map which implements such a choice: it chooses a
configuration of external momenta for the subgraph $\Gamma^\prime$,
permutes the orientation of external legs, or changes the insertion
place: any modification which will not modify the required
counterterm is allowed. By construction, elements in our ideal are
superficially convergent, and hence contribute nothing to the short
distance singular sector. The above ideal is determined by the
choice of the map ${\rm End}$, which is then determined by the
physics one wants to study.

We then decompose graphs as \be \Gamma=\underbrace{\sum_i
\Gamma^{\prime\prime}\circ_i {\rm End}^{-1}(\Gamma^\prime)
P(\Gamma^{\prime\prime})}_{p(\Gamma)} + \sum_i
\Gamma^{\prime\prime}\circ_i {\rm End}^{-1}\Gamma^\prime
P(\Gamma^{\prime\prime})P(\Gamma^{\prime}).\label{pG}\ee By
construction, the first term in the rhs is a primitive element
$p=p(\Gamma)$ in the Hopf algebra of decorated rooted trees, and
${\rm End}^{-1}={\rm End}\circ S$.

Note that for a primitive graph $\gamma$ and an endomorphism ${\rm End}$ which changes its external leg structure to say zero momentum transfer,
the above ideal simply says that the short distance singular sector of $\gamma$ and ${\rm End}(\gamma)$ are the same.

More instructive is the example of the case of say a graph with a
single divergent subgraph $\Gamma=\Gamma^{\prime\prime}\circ
\Gamma^{\prime}$. We then have \be
p(\Gamma)=\Gamma^{\prime\prime}\circ \left[\Gamma^{\prime}-{\rm
End}(\Gamma^{\prime})\right].\ee This is clearly primitive as
$\left[\Gamma^{\prime}-{\rm End}(\Gamma^{\prime})\right]$ is in the
chosen ideal, and $p(\Gamma)$ in (\ref{pG}) is primitive in general
because all its subgraph are in the ideal, by construction. We
emphasize that in all our applications we will never discard any
terms on the rhs of (\ref{pG}), but will always calculate the full
graph $\Gamma$, but use (\ref{pG}) as a convenient tool to come to a
manageable form for the primitives which drive the recursion, on the
expense to have an enlarged set of such integral kernels.

For our purposes, we choose an ideal which sets $m=0$ in structure
functions and evaluates external momenta at a symmetric point $-Q^2$
with regard to their external momenta, and iterates graphs into each
other in accordance with the definition of the Mellin transform
defined below. As a renormalization scheme we choose subtraction at
a fixed $-Q^2=\mu^2$.

\subsection{Mellin transform}
Any so constructed primitive $p=p(\Gamma)$ is a degree homogenous
combination of Hopf algebra elements of degree $|p|$. We let $-Q^2$
be the above kinematical variable which we keep. We define the
Mellin transform $M(p)$ of $p$ as \be M(p)(\rho)=[-Q^2]^\rho\int
{\rm
Int}_p(-Q^2)\left\{\frac{1}{|p|}\sum_{i=1}^{|p|}[k_i^2]^{-\rho}\right\}\prod_{i=1}^{|p|}d^4
k_i,\ee where ${\rm Int}_p(-Q^2)$ is the integrand determined by
$p$. We let \be{\rm Int}_p^-(-Q^2):={\rm Int}_p(-Q^2)-{\rm
Int}_p(\mu^2).\ee

We define renormalized Feynman rules for a symmetric momentum
scheme with subtractions at $-Q^2=\mu^2$ by \be
\phi_R(B_+^p(X))(-Q^2/\mu^2)=\int {\rm
Int}_p^-(-Q^2)\left\{\frac{1}{|p|}\sum_{i=1}^{|p|}\phi_R(X)(-k_i^2/\mu^2)\right\}\prod_{i=1}^{|p|}d^4
k_i.\ee

We have \be \phi_R(B_+^p(X))(-Q^2/\mu^2)=\lim_{\rho\to
0}\phi_R(X)(\partial_{-\rho})M(p)(\rho)\left[[-Q^2/\mu^2]^{-\rho}-1\right],\ee
where $\partial_{-\rho}=-\frac{\partial}{\partial \rho}$.

A Green function is then defined as the image under such Feynman
rules applied to a fixpoint of a combinatorial DSE. In the coming
sections, we first study the asymptotics of a combinatorial DSE,
then the growth after applying $\phi_R$.

Let us now look at an example in $\phi^4_4$ theory. We define the
vertex Green function by setting $-Q^2=s=t=u$, and set $m=0$ in all propagators in accordance with an ideal which
isolates the short distance singularities in massless Green functions.

Taking the symmetry in external legs into account, the Hopf algebra
series reads to order $g^2$ \be 1+g
\frac{3}{2}\paa+3g^2\left(\frac{1}{4}\pba+\frac{1}{2}\left[\pbb+\pbc\right]\right).\ee
We define a primitive \bea p_2 & = &
\frac{1}{4}\pba+\frac{1}{2}\left[\pbb+\pbc\right]-\frac{3}{2}\paa\circ{\rm
End}\left(\paa\right)\nonumber\\ & = &
\frac{1}{4}\pba-\frac{1}{2}\pbb.\eea

We define a combinatorial DSE \be
X(g)=1+g\frac{3}{2}B_+^{p_1}(X(g)^2)+3g^2B_+^{p_2}(X(g)^3),\ee
where $p_1=\paa$.

It
has a fixpoint which reproduces to order $g^2$ the above expansion.

Let us turn this into an integral equation: \bea G(g,\ln
-Q^2/\mu^2) & = & 1+\frac{3}{2}g\int \left\{\frac{G(g,\ln
-k^2/\mu^2
)^2}{k^2(k+Q)^2}\right\}_-d^4k\nonumber\\
 & + & 3g^2\int \frac{1}{8}\left[G(g,\ln -k^2/\mu^2)+G(g,\ln
   -l^2/\mu^2)\right]^3 \\
 & & \qquad \left\{
\frac{2k\cdot l-2(k+l)\cdot s-2s^2}{k^2l^2(k+Q)^2(l+Q)^2(l-k)^2}
 \right\}_-d^4k d^4l,\nonumber\eea where
$\}_-$ indicates subtraction at $-Q^2=\mu^2$. One verifies that the
solution to this integral equation agrees to order $g^2$ with the
perturbative renormalization in a symmetric momentum scheme for
the vertex function in $\phi^4_4$ theory, as it should. Its
solution also agrees to any order in $g^2$ with the leading order
in $L=\ln -Q^2/\mu^2$ in that order of $g^2$, and also with the
next to leading order, as we have taken the two loop primitive
into account.

Addition of further primitives delivers the lower powers in $L$. The
Mellin transform $M_{p_1}=\frac{1}{\rho(1-\rho)}$ is trivial, the
Mellin transform $M_{p_2}$ much less so and can be found in
\cite{BKW}.

\subsection{Mellin transforms as a geometric series}
The above Mellin transforms $M_p(\rho)$ have the form
\be M_p(\rho)=\frac{f_p(\rho)}{\rho(1-\rho)},\ee
in a natural manner, where the pole at $\rho=0$ reflects the short-distance singularity, and the pole at $\rho=1$ reflects
the fact that in our ideal we have massless internal propagation.

The series $f_p(\rho)={\rm res}_p+{\mathcal O}(\rho)$ determines the
residue ${\rm res}_p$ of the transform. We keep the factor
$1/(1-\rho)$ explicit though, to maintain the form of the Mellin
transform similar, in accordance with its conformal symmetries at
$-Q^2=1$.

By the above definition of the Feynman rules it is a purely
algebraic exercise \cite{BKW,dirkLL} to define a new series of
primitives $p_1=p$,
$p_2=B_+^p(B_+^p(\One))-\frac{1}{2}B_+^p(\One)B_+^p(\One)$, and so
on, such that \be M_p(\rho)=\sum_n q_n \frac{{\rm
res}_{p_n}}{\rho(1-\rho)}.\ee We henceforth assume a basis of
primitives $\{p\}$ such that \be M_p(\rho)=\frac{{\rm
res}_p}{\rho(1-\rho)},\ee which is convenient in the following. This
finishes the discussion of our analytic set-up and we next remind
ourselves of some basic properties of recursive systems relevant to
Dyson--Schwinger equations.

\section{The Universal law}
\subsection{The classical case}
A classical result of combinatorics is P\'olya's asymptotic
formula for the number, $t(n)$, of rooted, unlabelled trees with $n$
vertices \cite{po} (translated in \cite{pr}):
\begin{equation}\label{univlaw}
  t(n) \sim C \rho^{-n} n^{-3/2}
\end{equation}
where $C$ is an explicit constant and $\rho$ is the radius of
convergence of the generating function $\mathbf{T}(x) = \sum_{n\geq
  1}t(n)x^n$ of the
class of all rooted, unlabelled trees.  Thus $\rho$ is the reciprocal of
Otter's tree constant \cite{otter} (sequence A051491 in
\cite{OEIS}).  The key to P\'olya's proof is to convert the recursive
equation
\[
  \mathbf{T}(x) = x\exp\bigg(\sum_{m\geq 1}\mathbf{T}(x^m)/m\bigg)
\]
for rooted, unlabelled trees to a bivariate function
\[
  \mathbf{E}(x,y) = x e^y \exp\bigg(\sum_{m\geq 2}\mathbf{T}(x^m)/m\bigg)
\]
at which point the recursive equation becomes $\mathbf{T}(x) = \mathbf{E}(x,
\mathbf{T}(x))$.  Since $\rho < 1$ this rewriting uses the new
variable $y$ to isolate the
portion of the recursive equation which controls the radius of
convergence.

Consequently, $y - \mathbf{E}(x,y)$ is
amenable to Weierstrass preparation, which tells us that around
$(\rho, \mathbf{T}(\rho))$, $y-\mathbf{E}(x,y)$ is a product of a
holomorphic function which is nonzero around $(\rho,
\mathbf{T}(\rho))$ and a monic polynomial in $y$
with coefficients which are analytic in $x$.  The degree of the
polynomial is the order of the first nonzero derivative.
The failure of the implicit
function theorem at $\rho$ gives that
\begin{align*}
  \mathbf{T}(\rho) - \mathbf{E}(\rho,\mathbf{T}(\rho)) & = 0 \\
  1 - (\partial_y \mathbf{E})(\rho,\mathbf{T}(\rho)) & = 0 \\
  - (\partial_{yy}\mathbf{E})(\rho,\mathbf{T}(\rho)) & \neq 0.
\end{align*}
So the
Weierstrass polynomial is quadratic which gives that $\mathbf{T}$ has
a square root singularity at $\rho$:
\[
  \mathbf{T}(x) = f(x) + g(x)\sqrt{\rho-x}
\] with $f$ and $g$ analytic at $\rho$.  Then the
Cauchy integral theorem gives the desired asymptotics along with a
formula for $C$.

Asymptotics of the form $C \rho^{-n} n^{-3/2}$ have
since been widely found for classes of rooted trees
with recursive definitions and hence \eqref{univlaw} is
sometimes known as the universal law.
Some notable examples generalizing the reach of P\'olya's analysis
include \cite{can}, \cite{hrs}, and \cite{mm}.  Surveys include
\cite{drm} and \cite{odly}.  See also chapter VII of Flajolet
and Sedgewick's forthcoming book \emph{Analytic Combinatorics}
currently available as an online draft \cite{fs}\footnote{References
  are drawn from the draft of October 23, 2006.}.

For all but the last of the above the focus is on improved analytic conditions
which still must be checked for any particular case of interest.  One
approach for further generalization is to specify certain
combinatorial constructions and then use these constructions to build
recursive equations with solutions which automatically satisfy the
universal law.  This is the approach of \cite{univlaw}.
Flajolet and Sedgewick also use a framework of
combinatorial constructions to build their recursive equations,
getting the universal law for various schemas,
such as the exp-log schema \cite[VII.1]{fs}.

\subsection{Recursive systems}
Another approach to generalizing P\'olya's analysis is to move to
recursive systems of equations while restricting the complexity of
each equation.  This is of particular relevance to
applications to counting Feynman diagrams since generally only
polynomials and geometric series are needed,
but systems are difficult to avoid.

First note that any geometric series
can be converted to a
polynomial at the expense of a new variable.
Namely replace $1/(1-X)$
with a new variable $F$ and add the equation
\[
  F = 1 + F \cdot X.
\]
In view of this reduction we can focus on polynomial systems.

Nonnegative polynomial recursive systems give the universal law under
some reasonable conditions, as was shown independently by
Drmota\cite{drsys}, Lalley\cite{lal}, and Woods\cite{wo}.  For our
purposes the full generality of the above are not necessary and we'll
follow the presentation of Flajolet and Sedgewick \cite[VII.6.3]{fs}.

Suppose
\begin{align*}
  y_1 & = \Phi_1(x, y_1, \ldots, y_m) \\
  & \vdots \\
  y_m & = \Phi_m(x, y_1, \ldots, y_m)
\end{align*}
with the $\Phi_i$ polynomials with real coefficients.

There are five conditions which together guarantee that each component
solution to this system satisfies the universal law.  First we say the
system is \emph{nonlinear} if at least one of the $\Phi_i$ is
nonlinear in $y_1, \ldots, y_m$.  The system is \emph{nonnegative} if
each $\Phi_i$ has nonnegative coefficients.

The next condition guarantees that the system does in fact behave
recursively.  For
$\overline{y} = (y_1, \ldots, y_m) \in \mathbb{R}[[x]]^m$ define the
$x$-valuation by $\text{val}(\overline{y}) = \min_i (\text{val}(y_i))$
where the valuation of a series picks out the index of the first
nonzero coefficient, i.e. $\text{val}(\sum_{n=k}^\infty a_n x^n) = k$
when $a_k \neq 0$, and with the convention that $\text{val}(0) =
\infty$.  Also define
$d(\overline{y}, \overline{y}') = 2^{-\text{val}(\overline{y} -
  \overline{y}')}$.  Then the system is \emph{proper} if
\[
d(\Phi(\overline{y}), \Phi(\overline{y}')) < K d(\overline{y}, \overline{y}') \qquad \text{for some $K<1$}.
\]

To guarantee that the system behaves as one system rather than many
we need to guarantee that all variables play a role in each
equation.  Specifically, define the dependency graph of the system to
be the directed graph on $\{1, \ldots, m\}$ with an edge from $k$ to
$j$ if $y_j$ figures in a monomial of $\Phi_k(x, \overline{y})$.
The system is \emph{irreducible} if the dependency graph is strongly
connected, that is there is a path between any two vertices following
edges only in their forward direction.

Finally, to remove spurious zero coefficients in the solution series,
and to avoid extra singularities on the circle of convergence, define
a power series $\mathbf{T}(x)$ to be \emph{aperiodic} if
it is not the case that there is a power series $\mathbf{U}(x)$ and
integers $a\geq 0$ and $d \geq 2$ such that $\mathbf{T}(x) =
x^a\mathbf{U}(x^d)$. If the solutions for each $y_i$ of the system
are aperiodic then the system itself is said to be \emph{aperiodic}.

Then
\begin{thm}[\cite{fs} Theorem VII.6]\label{systems result}
  Suppose $\overline{y} = \Phi(\overline{y})$ is a polynomial system
  that is nonlinear, proper, nonnegative, and irreducible.  Then all
  component solutions $y_j$ have the same radius of convergence $\rho
  < \infty$ and have a square root singularity at $\rho$.
  If furthermore the system is aperiodic then all $y_j$ satisfy the
  universal law.
\end{thm}

An outline of the proof is as follows.  For more details see
\cite[Theorem VII.6]{fs}.  Properness gives that the
component solutions are unique, because the solutions can be generated recursively from
the zero vector, and along with nonnegativity we get
further that the component solutions have nonnegative coefficients.
Irreducibility, along with nonnegativity to prevent cancellations,
gives that the radius of each $y_i$ is the same value $\rho$.
Aperiodicity forces $\rho$ to be the only singularity of each
component solution on the circle of convergence.

The key to the proof is the square root singularity at $\rho$.  Each
of the three independent proofs takes a different approach, though in
all cases, as in the single equation situation, it comes down to using the
failure of the implicit function theorem at $\rho$. Nonlinearity is
necessary to avoid singularities which themselves are linear rather
than of square root type.

Drmota \cite{drsys} proceeds by solving a subset of the equations for
the remaining variables and then substituting back into the remaining
equations to reduce the number of equations in the system.
Iteratively he is able to reduce to one equation at which point the
system can be treated classically.

Lalley \cite{lal}, summarized in \cite[Theorem VII.6]{fs}, considers the linearized
Jacobian
\[
  \bigg(\partial_{y_j} \Phi_i(x, y_1, \cdots y_m)\bigg)
\]
and uses Perron-Frobenius theory to show that the largest eigenvalue
in absolute value at $x=\rho$
of the Jacobian is precisely $1$ and that there is an eigenvector $v$
with positive coefficients.  Then multiplying the system by $v$ and
expanding around $\rho$ gives the desired asymptotics.

Woods \cite{wo} also uses the Jacobian and the largest eigenvalue $1$.  He
continues the analysis on block upper triangular matrices in order to
deal with certain non-irreducible cases.

\section{The universal law for Feynman diagrams}

\subsection{QED with one primitive per loop order}
Massless quantum electrodynamics provides three monomials in the Lagrangian which need renormalization,
corresponding as one-particle irreducible Green functions to the inverse fermion and photon propagators,
and the vertex. We do not yet impose the Ward identity and proceed by setting up the combinatorial structure as it rather typically holds
for a theory with three-valent vertices and two types of edges (the degenerate case of a single type of edge is easily interfered).

The smallest Hopf algebra which allows for a correct renormalization of the full theory treats the sum of all primitive graphs at a given loop order as
a single Hochschild cocycle, and we start from there.
%
%So consider the example of QED with one primitive per loop order.  The
%single primitive can be interpreted as the sum of all connected
%primitive graphs which gives a canonical and important sub Hopf
%algebra.
This situation is
described by the following system.
\begin{align*}
  X_1 & = 1 + \sum_{k \geq 1} x^k \frac{X_1^{2k+1}}{(1-X_2)^{2k}(1-X_3)^
k} \\
  X_2 & = x \frac{X_1}{(1-X_2)(1-X_3)} \\
  X_3 & = x \frac{X_1}{(1-X_2)^2}
\end{align*}
%Its usual graph version: \input{vertexdse.tex}
%\input{fermiondse.tex}
%\input{photondse.tex}

After converting the geometric series we get
\[
  \Phi =
  \begin{cases}
    X_1 & = 1 + X_1 F_1 \\
    F_1 & = xX_1^2F_2^2F_3 + x X_1^2F_2^2F_3F  \\
    X_2 & = x X_1F_2F_3 \\
    F_2 & = 1 + F_2X_2 \\
    X_3 & = x X_1F_2^2 \\
    F_3 & = 1 + F_3X_3
  \end{cases}
\]
$\Phi$ is nonnegative, polynomial, nonlinear, and irreducible.  $\Phi$
can be seen to be aperiodic simply by calculating the first few
terms.  $\Phi$ itself is not proper.  However $\Phi^2$
describes the same solutions and the other properties, nonnegative,
polynomial, nonlinear, irreducible, and aperiodic, remain true for
powers.

To see that $\Phi^2$ is
proper suppose we have 2 vectors
\[
  v = (x_1, f_1, x_2, f_2, x_3, f_3)
\]
and
\[
  v' = (x_1', f_1', x_2', f_2', x_3', f_3')
\] at distance $2^{-n}$; so
write $x_1' = x_1 + x_1''$ with $x_1''$ having no term of degree less
than $n$ and
similarly for the other coordinates.
Consider the difference in $X_1$ coordinates after applying $\Phi$:
\[
  x_1f_1 - x_1'f_1' = - x_1''f_1'' - x_1''f_1 - x_1f_1''
\]
which has no terms of
degree less than $n$.  Further if $f_1$ has no constant term and $f_1''$
has no term of degree $n$ then the difference in the $X_1$ coordinate
has no terms of degree less than $n+1$.
Argue similarly for $F_2$ and $F_3$.
 For the $X_2$ coordinate after applying $\Phi$ we get a difference of:
\[
xx_1f_2f_3 - xx_1'f_2'f_3' = -xx_1''f_2''f_3'' - xx_1''f_2''f_3 -
xx_1''f_2f_3'' - xx_1f_2''f_3'' - xx_1f_2''f_3 - xx_1f_2f_3''
\]
which
has no terms of degree less than $n+1$.  Notice also that the new
$X_2$ coordinate has no constant term for both initial vectors.  Argue similarly for $X_3$ and $F_1$.

Now consider applying $\Phi$ a second time.  Apply the above arguments
again but notice that we are in the ``further'' case for $X_1$, $F_2$
and $F_3$, so all coordinates now have a difference with no terms of
degree less than $n+1$, that is the distance has decreased by at least
$1/2$.  Thus any $1 > K > 1/2$ will give that $\Phi^2$ is proper.

Consequently by Theorem \ref{systems result} we know that all 6 solution series to $\Phi^2$ and hence all
3 solution series to the original system have the same radius of
convergence $\rho$ and have coefficients with the asymptotic form
\[
  C \rho^{-n} n^{-3/2}
\]

In order to understand the asymptotic growth rate for the QED system
with primitives summed at each loop order it remains to understand the
radius $\rho$. Substituting $X_2$ and $X_3$ into $F_2$ and $F_3$
respectively we get $F_2 = 1 + xX_1F_2^2F_3 = F_3$.  Thus the original
system can be rewritten
\begin{align*}
  X_1 & = 1 + \frac{xX_1^3F_2^3}{1 - xX_1^2F_2^3} \\
  F_2 & = 1 + xX_1F_2^3
\end{align*}
Rearrange
\[
  X_1
  = 1 + \frac{X_1^2(F_2-1)}{1-X_1(F_2-1)}
\]
then solve to get
\[
  F_2 = \frac{1 - 2X_1^2}{X_1(1 - 2X_1)}
\]
Substitute back into the equation for $F_2$ and expand to get
\[
  -x + X_1 + (6x-5)X_1^2 + 8X_1^3 + (-12x-4)X_1^4 + 8xX_1^6 = 0
\]
As a polynomial in $X_1$ this has discriminant
\[
  4096 x^2  (32 x^2  - 8 x + 1) (-2 + 27 x)^2
\]
So the radius of the system is
\[
  \frac{2}{27}
\]

In view of the fact that the radius is an important value associated
to the system and that this system is canonically associated to QED
we're led to the following question as to what is the physical
meaning of $2/27$ in QED? The relevance of this number extends to
any system which provides a recursive system similar to QED. We can
proceed similarly for other theories; the results of some examples
are summarized in Appendix \ref{others}.

\subsection{Polynomially many primitives per loop order}
Combining all primitives at a given loop order into one Hochschild
cocycle which drives the Dyson--Schwinger equation defines the
smallest sub Hopf algebra which still renormalizes the full theory
correctly. It is often instructive to disentangle the primitives in
different ways, for example in accordance with the transcendental
nature of ${\rm res}_p$. This motivates to consider a slightly more
general condition on the number of primitives, enough to apply the
polynomial systems result. we next assume there to be  $p(k)$
primitives at $k$ loops where $p$ is a polynomial.

 To see that this circumstance
reduces to a nonnegative polynomial system it suffices to show that
\[
  \sum_{k \geq \ell}p(k)B^k
\]
can be written as a sum of powers of geometric series.  This follows
from two
facts.

First, the falling factorials
\[
  \{ k(k-1)\cdots (k-n+1) : n \geq 0 \}
\]
form a basis for polynomials in $k$, and
\[
  k^n = \sum_{j=1}^n  S_2(n,j) k(k-1)\cdots (k-j+1)
\]
where $S_2(n,j)$ are the Stirling numbers of the second kind, A008277
in \cite{OEIS}, so in particular are nonnegative, and hence
for $p(k)$ with nonnegative coefficients we only need nonnegative
coefficients of the falling factorials.
%The Stirling number identity
%is a standard result which can be seen from the generating function
%\begin{align*}
%  \sum_{n} S_2(n,j)\frac{x^n}{n!} & = \frac{(e^x-1)^j}{j!} \\
%  \intertext{which gives}
%  \sum_{n,j} S_2(n,j)k(k-1)\cdots (k-j+1)\frac{x^n}{n!} & = \sum_{j
%    \geq 0}\binom{k}{j}(e^x-1)^j \\
%  & = e^{xk}
%  \intertext{so}
%  \sum_{j} S_2(n,j) k(k-1) \cdots (k-j+1) & = k^n
%\end{align*}

Second, notice that for $\ell \geq n$
\begin{align*}
  & \sum_{k \geq \ell} k(k-1) \cdots (k-n+1) B^k \\
  & = B^n \frac{d^{n}}{d B^{n}} \sum_{k \geq \ell} B^{k} \\
  %& = B^n \frac{d^n}{d B^n} \frac{B^{\ell}}{1-B} \\
  & = B^n \sum_{j = 0}^{n}\binom{n}{j} \left(\frac{d^j}{ d B^j} \frac{1}{1-B}\right)\left(
  \frac{d^{n-j}}{d B^{n-j}} B^{\ell} \right)\\
  & = \sum_{j=0}^{n} \ell(\ell-1) \cdots (\ell-n+j+1) \binom{n}{j}
  \left(\frac{1}{1-B}\right)^{j+1} B^{\ell+j}
\end{align*}
where all the coefficients are nonnegative.

%With polynomially many primitives per loop order the interpretation is
%not in general as natural as for 1 primitive per loop order.  We can
%view the primitives as broken into $p(k)$
%blocks and consider the system in terms of the blocks; for small $k$
%we can either over count or set up finitely many leading terms of the
%system separately.

One case where there is a natural interpretation is QED with a linear
number of generators, namely
\[
  X_1 = 1 + \sum_{k \geq 1} p(k)x^k
  \frac{X_1^{2k+1}}{(1-X_2)^{2k}(1-X_3)^k}
\]
with $X_2$ and $X_3$ as before and with $p(k)$ linear,
which corresponds to counting with Cvitanovi\'c's gauge
invariant sectors \cite{cvitanovic}.

%***we can leave this bit out if you like, I just thought it better to
%start with it in***
%We can calculate the radius numerically for QED with with $p(k) =
%k^\ell$ and graph as a function of $\ell$, see figure \ref{radiigraph}.
%The resulting graph has an
%inflection point around $\ell = 8$ which leads us to ask
%\begin{question}
%  Are degree $8$ polynomials of particular interest for approximating
%  QED by using a polynomial number of generators per loop order?  Can
%  we understand more generally the nature and importance of the radius
%  as a function of the degree of the monomial?
%\end{question}
%
%\begin{figure}
%\epsfig{file=radiioutput.ps, scale=0.5, angle=-90}
%\caption{\label{radiigraph}***I'll make a cleaner version which doesn't have 4 points per
%  point for obscure reasons ***}
%\end{figure}

\subsection{Other systems}
Johnson, Baker, and Willey \cite{jbw} use gauge invariance to reduce
the QED system to
\[
  X = x \sum_{k\geq 0} \left(\frac{x}{1-X}\right)^k = \frac{x(1-X)}{1-X-x}
\]
While amenable to the universal law analysis, this recursive
equation can be solved exactly by the quadratic formula.  We get
\[
 X = \frac{1 + \sqrt{1 -4x}}{2}
\]
giving the Catalan numbers, A000108 in \cite{OEIS}, as coefficients.
The radius
is $1/4$ which is considerably larger than $2/27$, showing how powerful gauge invariance is.
Note that it is only the inverse photon propagator $1-X$ which needs renormalization, and that it appears in the denominator.

\section{The growth of $\gamma_1$}
After these considerations of the combinatorial side, we discuss analytic aspects.
\subsection{The recursions}%\subsection{Uniformity}
Consider the Dyson--Schwinger equation
\[
  X(x) = \mathbb{I} - \sum_{k\geq 1}\sum_{i=0}^{s_k}x^k
  p_i(k) B^{k,i}_+(X Q^{k})
\]
where $Q = X^r$ with $r < 0$ an integer, and $p_i(k)$ coefficients,
not necessarily polynomial.
We'll use the notation
$
  F_{k,i}(\rho)% = \frac{1}{\rho}\sum_{j \geq 0} m_{j,k,i} \rho^j
$
for the Mellin transform of the integral kernel. Starting with $r<0$ is justified in light of the last example of the previous section, we will generalize this soon enough.

Specializing \cite[(26)]{non lin} to this case we get the recursion
\begin{equation}\label{higher recursion}
  \gamma_k(x) = \frac{1}{k} \gamma_1(x)(1+rx \partial_x)\gamma_{k-1}(x)
\end{equation}
independently of the $p_i(k)$.
Using the dot notation, $\gamma \cdot U = \sum \gamma_k U^k$, of \cite{non lin} we have
\begin{equation}\label{dot DSE}
  \gamma \cdot L = \sum_k \sum_ix^k p_i(k) (1 + \gamma \cdot
  \partial_{-\rho})^{-rk+1}(1-e^{-L\rho})F_{k,i}(\rho) |_{\rho=0}
\end{equation}
Taking one $L$ derivative and setting $L$ to $0$ we get
\begin{equation}\label{1 recursion}
  \gamma_1 = \sum_k \sum_i x^k p_i(k) (1 + \gamma \cdot
  \partial_{-\rho})^{-rk+1} \rho F_{k,i}(\rho) |_{\rho=0}
\end{equation}

%{}From the above we can derive a uniformity result in the $m_{j,k,i}$.
%\begin{lemma}
%  Writing $\gamma_{k, j} = \sum c_{k, j, j_1, \cdots j_u,
%    \overline{\ell}, \overline{i}} m_{j_1,
%    \ell_1, i_1} \cdots m_{j_u, \ell_u, i_u}$
%  for $j \geq k$ we have that $j_1 + \cdots + j_u = j - k$
%\end{lemma}
%
%\begin{proof}
%  The proof proceeds by induction.  Call $j_1 + \cdots + j_u$ the
%  $m$-degree of $\gamma_{k,j}$.  First note that $\gamma_{1,1} =
%  \sum_i m_{0,1,i} p_i(1)$ from (\ref{1 recursion}).
%
%  Assume the result holds for $k, j < n$.
%
%  Then from (\ref{1 recursion}) $\gamma_{1, n}$ is a
%  sum over $s$ of terms of the form
%  \begin{equation}\label{gamma1n piece}
%    C \gamma_{\ell_1, t_1} \cdots \gamma_{\ell_u, s_u} m_s
%  \end{equation}
%  where $\ell_1+ \cdots + \ell_u = s$ and $t_1 + \cdots t_u = n-1$.
%  By the induction hypothesis $\gamma_{\ell_i, t_i}$ has $m$-degree
%  $t_i - \ell_i$, so (\ref{gamma1n piece}) has $m$-degree $\sum t_i
%  - \sum \ell_i + s = n-1 -s +s = n-1$ as desired.
%
%  Next from (\ref{higher recursion}) $\gamma_{k, j}$, for $k,j \leq n$, is a
%  sum over $1 \leq i \leq n$ of terms of the form
%  \begin{equation}\label{gammanj piece}
%    C \gamma_{1, i} \gamma_{k-1, j-i}
%  \end{equation}
%  By the induction hypothesis $\gamma_{1, i}$ has $m$-degree $i-1$ and
%  $\gamma_{k-1, j-1}$ has $m$-degree $j-i-k+1$ so (\ref{gammanj
%    piece}) has $m$-degree $j-k$ as desired.
%\end{proof}

%\subsection{Finding the radius}
Restricting to $\rho F_{k,i}(\rho) = r_{k,i}/(1-\rho)$ allows us to write a much
tidier recursion for $\gamma_1$.
Taking two $L$ derivatives of (\ref{dot DSE}) and setting $L=0$ we get
\begin{align*}
  2 \gamma_2 & = - \sum_k \sum_i x^k p_i(k) (1 + \gamma \cdot
  \partial_{-\rho})^{-rk+1} \rho^2 F_{k,i}(\rho) |_{\rho =0} \\
  & = - \sum_k \sum_i x^k p_i(k) (1 + \gamma \cdot
  \partial_{-\rho})^{-rk+1}
  \rho F_{k,i}(\rho)|_{\rho = 0} + \sum_k x^k r_{k,i}p_i(k) \\
  & = - \gamma_1 + \sum_{k\geq 1}\sum_i r_{k,i}p_i(k) x^k \qquad \text{from
    (\ref{1 recursion})}
\end{align*}
Thus we can from now on ignore the sum over $i$ and let $p(k) =
\sum_{i}r_{k,i}p_i(k)$.  Then from (\ref{higher recursion})
\[
 \gamma_1 = \sum_{k\geq 1} p(k) x^k - 2 \gamma_2 = \sum_{k\geq 1} p(k) x^k - \gamma_1(1+rx\partial_x)\gamma_1
\]
giving
\begin{prop}
\[
  \gamma_{1,n} = p(n) + \sum_{j=1}^{n-1} (-rj-1)\gamma_{1,j}\gamma_{1,n-j}
\]
\end{prop}

\subsection{Finding the radius}
We see from the proposition that if $\sum p(k)x^k$ is
Gevrey-$n$, that is $\sum x^k p(k)/(k!)^n$ converges, but not
Gevrey-$(n-1)$, then
$\gamma_1$ is at best Gevrey-$n$.

Of most interest for our applications are the cases
where only finitely many $p(k)$ are nonzero but all are nonnegative
and where $p(k) = c^k k!$
giving the Lipatov bound; so $\sum p(k)x^k$ is Gevrey-1.  Assume that $p(k) \geq 0$ and
\[
   \sum_{k \geq 1} x^k \frac{p(k)}{k!} = f(x)
\]
has radius $0 < \rho \leq \infty$ and $f(x) > 0$ for $|x|\leq \rho$.    The
above two cases are included as $f(x)$ a polynomial
and $f(x) = cx/(1-cx)$ respectively.
In such circumstances $\gamma_1$ is also Gevrey-1 and the radius is the minimum of $\rho$ and
$-1/(ra_1)$ (where we view $-1/(ra_1)$ as $+ \infty$ in the case $a_1
= 0$) which we can see as follows\footnote{Similarly $\sum p(k)x^k$
  Gevrey-$n$ for $n > 1$ leads to $\gamma_n$ Gevrey-$n$}.

Let $a_n = \gamma_{1,n}/n!$. Then $a_1 = \gamma_{1,1} = p(1)$ and
\begin{align*}
  a_n = & \frac{p(n)}{n!} + \sum_{j=1}^{n-1} (-rj-1)\binom{n}{j}^{-1}
  a_j a_{n-j} \\
  = & \frac{p(n)}{n!} + \frac{1}{2}\sum_{j=1}^{n-1} (-rj-1
  -r(n-j)-1)\binom{n}{j}^{-1} a_j a_{n-j} \\
  = & \frac{p(n)}{n!} + \left(-r\frac{n}{2} -1\right)\sum_{j=1}^{n-1} \binom{n}{j}^{-1}
  a_j a_{n-j}
\end{align*}

To achieve an upper bound on the radius of convergence of $\sum a_n
x^n$ take the first and last terms of the sum to get
\[
  a_n \geq \frac{p(n)}{n!} -r\frac{n-2}{n}a_1a_{n-1}
\]
for $n \geq 2$.
So the radius of $\sum a_n x^n$ is no more than the radius of the
recursively defined series with equality above, say
\[
  b_n = \frac{p(n)}{n!} -r\frac{n-2}{n}b_1b_{n-1}
\]
for $n \geq 2$ with $b_1= a_1$.  Immediately we see that if $a_1 = 0$
the radius of $\sum b(n) x^n$ is $\rho$.  Otherwise consider
\[
  n(n-1)b_n = \frac{p(n)n(n-1)}{n!} - r (n-1)(n-2)b_1b_{n-1}
\]
Equivalently with $\mathbf{B}(x) = \sum b(n) x^n$ we get
\[
  \mathbf{B}''(x) = f''(x) - rb_1x \mathbf{B}''(x)
\]
Solving for $\mathbf{B}''(x)$
\[
  \mathbf{B}''(x) = \frac{f''(x)}{1+ra_1x}
\]
which, since differentiation does not change the radius of a series,
has radius $\min \{\rho, -1/(ra_1)\}$, and thus so does $\mathbf{B}(x)$.

For the lower bound on the radius we need a few preliminary
results.  First a simple combinatorial fact.

\begin{lemma}
  Given $0 < \theta < 1$
  \[
    \frac{1}{n} \binom{n}{j} \geq \frac{\theta^{-j+1}}{j}
  \]
  for $1 \leq j \leq \theta n$ and $n \geq 2$.
\end{lemma}

\begin{proof}
  Fix $n$.  Write $j = \lambda n$, $0 < \lambda \leq \theta$.  Then
  \[
    \frac{1}{n} \binom{n}{j} =
    \frac{1}{n} \binom{n}{\lambda n} \geq  \frac{n^{\lambda
        n-1}}{(\lambda n)^{\lambda n}}
    = \frac{\lambda^{-\lambda n + 1}}{\lambda n} \geq
    \frac{\theta^{-j+1}}{j}
  \]
\end{proof}

Second we need to understand the behaviour of $\sum a_n x^n$ at the
radius of convergence.

\begin{lemma}\label{converge}
  Using notation as above and with $\mathbf{A}(x) = \sum a_n x^n$ with
  radius of convergence
  $\rho_a$ we
  have that $\limsup_{x \rightarrow \rho_a}\mathbf{A}(x)(1+xra_1)/f(x)
  \leq 1$
\end{lemma}

\begin{proof}
  Take any $0 < \theta < 1/2$.
  Using the previous lemma
  \begin{align*}
    a_n = & \frac{p(n)}{n!} + \left(-r\frac{n}{2}
      -1\right)\sum_{j=1}^{n-1} \binom{n}{j}^{-1}  a_j a_{n-j} \\
    \leq & \frac{p(n)}{n!} -rn\sum_{1
      \leq j \leq \theta n}  \binom{n}{j}^{-1} a_j a_{n-j} -r
      \frac{n}{2} \sum_{\theta n \leq j \leq n-\theta
      n}\binom{n}{j}^{-1} a_j a_{n-j} \\
    \leq & \frac{p(n)}{n!} -r \sum_{1
      \leq j \leq \theta n} j\theta^{j-1} a_j a_{n-j}
    -r \binom{n}{ \theta n }^{-1}\frac{n}{2}
      \sum_{\theta n \leq j \leq n-\theta
      n}a_j a_{n-j} \\
    \leq & \frac{p(n)}{n!} - r \sum_{1
      \leq j \leq \theta n} j \theta^{j-1} a_j a_{n-j}
    -\frac{r}{2} n \theta^{\theta n}
    \sum_{\theta n \leq j \leq n-\theta
      n} a_j a_{n-j}
  \end{align*}
  so the coefficients of $\mathbf{A}(x)$ are bounded above by the
  coefficients of
  \[
    f(x) - xr\mathbf{A}'(\theta x) \mathbf{A}(x) -
    \frac{r}{2}\left(x\frac{d}{dx}\mathbf{A}^2\right)( \theta^\theta
    x)
  \]
  Since all coefficients are nonnegative, for any $0 < x <
  \rho_a$ we have
  \[
    \mathbf{A}(x) \leq f(x) - xr\mathbf{A}'(\theta x) \mathbf{A}(x) -
    \frac{r}{2}\left(x\frac{d}{dx}\mathbf{A}^2\right)( \theta^\theta
    x)
  \]
  which is continuous in $\theta$, so for fixed $0 < x <
  \rho_a$ we can let $\theta \rightarrow 0$ giving
  \[
    \mathbf{A}(x) \leq f(x) - xra_1\mathbf{A}(x)
  \]
  so
  \[
    \frac{\mathbf{A}(x)(1+xra_1)}{f(x)} \leq 1
  \]
  for $0 < x < \rho_a$.  The result follows.
\end{proof}

Let $\rho_a$ be the radius of $\sum a_n x^n$.
If $\rho_a = \rho$ then we're done, so suppose $\rho_a < \rho$.  To
get the lower bound on $\rho_a$ it
remains then to prove that $\rho_a \geq  -1/(ra_1)$ when $a_1 \neq 0$
and to prove a contradiction in the case that $a_1 = 0$.
Take any
$\epsilon > 0$.  Then there
exists an $N>0$ such that for $n>N$
\begin{align*}
 a_n & \leq \frac{p(n)}{n!} - ra_1a_{n-1} - r\frac{1}{n-1}\sum_{j =
   2}^{n-2}a_ja_{n-j}\\
 & \leq \frac{p(n)}{n!} - ra_1a_{n-1} + \epsilon \sum_{j=2}^{n-2}a_j a_{n-j} \\
 & \leq \frac{p(n)}{n!} - ra_1a_{n-1} + \epsilon \sum_{j=1}^{n-1}a_ja_{n-j}
\end{align*}

Define
\[
  c_n = \begin{cases}
    a_n & \text{if } a_n > \frac{p(n)}{n!} - rc_1c_{n-1} +
    \epsilon \sum_{j=1}^{n-1}c_jc_{n-j} \\
    \frac{p(n)}{n!} - rc_1c_{n-1} + \epsilon
    \sum_{j=1}^{n-1}c_jc_{n-j} & \text{otherwise (in particular
      when $n>N$)}
    \end{cases}
\]
In particular $c_1 = a_1$.
The radius of $\sum a_n x^n$ is at least as large as the radius
of $\mathbf{C}(x) = \sum c_n x^n$.  Rewriting with generating series
\[
  \mathbf{C}(x) = f(x) -ra_1x\mathbf{C}(x) + \epsilon
  \mathbf{C}^2(x) + P_\epsilon(x)
\]
where $P_\epsilon(x)$ is some polynomial.  This equation can be solved by the
quadratic formula.  The discriminant is
\begin{equation}\label{discriminant}
  (ra_1x + 1)^2 - 4 \epsilon (f(x)+P_\epsilon(x))
\end{equation}
By assumption we are interested in $|x| < \rho$ where (\ref{discriminant})
has no singularities, so
the radius of $\mathbf{C}(x)$ is the closest root to
$0$ of (\ref{discriminant}); call it $\rho_\epsilon$.  Consider
$\epsilon = 0$ giving
\[
  (ra_1 x + 1)^2
\]
which has $-1/(ra_1)$ as its closest root to $0$ when $a_1 \neq 0$; call
this value $\rho_0$.

Then to get the lower bound on the radius of $\sum a_n x^n$ it remains
only to prove the following lemma.

\begin{lemma}
  With notation and assumptions as above if $a_1 \neq 0$, $\lim_{\epsilon \rightarrow 0}
  |\rho_\epsilon| \geq \rho_0$, while if $a_1 = 0$ we have a contradiction.
\end{lemma}

\begin{proof}
  In view of Lemma \ref{converge} this is a short exercise in analysis.

  By construction the coefficient of $x^n$ in $P_\epsilon(x)$ is
  bounded by $a_n + rc_1c_{n-1} \leq a_n + ra_1a_{n-1}$ since $c_n
  \geq a_n$ for all $n \geq 1$.  So $P_\epsilon(x)$ has coefficients
  which are nonnegative and bounded by those of $\mathbf{A}(x)(1+rxa_1)$.  Thus
  by Lemma \ref{converge}, the continuity of $P_\epsilon(x)$ at $\rho_a$, and the assumption that $\rho_a < \rho$, we
  see that $f(\rho_a) + P_\epsilon(\rho_a) \leq f(\rho_a) + \liminf_{x
    \rightarrow \rho_a}(\mathbf{A}(x)(1+rxa_1)) < \infty$.   By the
  nonnegativity of the
  coefficients of $f$ and $P_\epsilon$ we can choose $M>0$ such that $|f(x) +
  P_\epsilon(x)| < M$ independently of $\epsilon$ for $|x| \leq
  \rho_a$.

  Suppose $a_1 \neq 0$.  Take any $\eta > 0$.   Consider $|x| \leq \rho_a$.
  Choose $\delta > 0$ such that
  $(ra_1x+1)^2 < \delta$ implies $|x - \rho_0| < \eta$.
  Pick $\epsilon < \delta/(4M)$.  Then
  $
    (ra_1\rho_\epsilon +1)^2 = 4\epsilon(f(\rho_\epsilon) +
    P_\epsilon(\rho_\epsilon)) < \delta
  $
  so $|\rho_\epsilon - \rho_0| < \eta$

  Suppose on the other hand that $a_1 = 0$.  Take $0 < \delta < 1$.
  Then, since $|\rho_\epsilon|
  \leq \rho_a$, we get that for $\epsilon < \delta/(4M)$, $1 =
  4\epsilon(f(\rho_\epsilon) + P_\epsilon(\rho_\epsilon)) < \delta$
  which is a contradiction.
\end{proof}

Taking the two bounds together we get the final result

\begin{thm}
Assume $\sum_{k \geq 1} x^k p(k)/k!$ has radius $\rho$.
Then $\sum x^n\gamma_{1,n}/n!$ converges with radius
of convergence $\min \{\rho, -1/(r\gamma_{1,1})\}$, where
$-1/(r\gamma_{1,1})$ is
interpreted to mean $+\infty$ in the case $\gamma_{1,1} = 0$.
\end{thm}
\subsection{Nonnegative systems}

Now suppose we have a system of Dyson-Schwinger equations
\[
  X^r(x) = \mathbb{I} - \sum_{k \geq 1}\sum_{i=0}^{s_k}x^k
  p_i^r(k) B_+^{k,i;r}(X^rQ^{k})
\]
for $r \in \mathcal{R}$ with $\mathcal{R}$ a finite set, $p_i^r(k) \geq
0$, and
where
\[
  Q = \prod_{r\in \mathcal{R}}X^r(x)^{s_r}
\]
with integers $s_r < 0$ for all $r \in \mathcal{R}$.
%\begin{equation}\label{p assumption}
%  \sum_{r \in \mathcal{R}} -s_r p^r(n) \geq p^t(n)
%\end{equation}
%for all $t \in \mathcal{R}$ and all $n \geq 1$ and with
%\begin{equation}\label{base assumption}
%  \sum_{r \in \mathcal{R}} -s_r \gamma_{1,1}^r \geq \gamma_{1,1}^t
%\end{equation}
%for all $t \in \mathcal{R}$.

%The assumptions (\ref{p assumption}) and (\ref{base assumption}) guarantee
%that the $\gamma_{1,1}^r$ have nonnegative coefficients knowing the
%recursions for the $\gamma_{1}^r$.

Then as before from \cite[(26)]{non lin} we have
\begin{equation}\label{systems higher recursion}
  \gamma^r_k(x) = \frac{1}{k}\left( \gamma^r_1(x)+ \sum_{j
      \in \mathcal{R}}s_j\gamma^j_1(x)x \partial_x\right)\gamma^r_{k-1}(x)
\end{equation}
again independent of the $p^r(k)$.

Assume that there is one insertion place, and so one variable $\rho$,
and that the Mellin transform of the integral kernel is a geometric
series $\rho F^r_{k,i}(\rho) = r_{k,i;r}/(1-\rho)$.

Rewriting the system of Dyson-Schwinger equations with dot notation we
have
\begin{equation}\label{systems dot DSE}
  \gamma^r \cdot L = \sum_k \sum_i x^k p_i^r(k) \prod_{j \in
    \mathcal{R}}(1 + \gamma^j \cdot \partial_{-\rho})^{-s_jk+1}(1-e^{-L\rho})F^r_{k,i}(\rho) |_{\rho=0}
\end{equation}
As before we can find tidier recursions for the $\gamma^r_1$ by
comparing the first and second $L$ derivatives of (\ref{systems dot
  DSE}).  We get
\[
  \gamma^r_1 = \sum_k \sum_i x^k p_i^r(k)\prod_{j \in
    \mathcal{R}}(1 + \gamma^j \cdot \partial_{-\rho})^{-s_jk+1}
  \rho F^r_{k,i}(\rho) |_{\rho=0}
\]
and
\begin{align*}
  2 \gamma^r_2 & = - \sum_k \sum_i x^k p_i^r(k) \prod_{j \in
    \mathcal{R}}(1 + \gamma^j \cdot \partial_{-\rho})^{-s_jk+1} \rho^2
  F^r_{k,i}(\rho) |_{\rho =0} \\
  & = - \gamma^r_1 + \sum_{k\geq 1}\sum_i r_{k,i;r}p_i^r(k) x^k \qquad \text{since } \rho F_{k,i}(\rho) = \frac{r_{k,i;r}}{1-\rho}
\end{align*}
Thus letting $p(k) = \sum_i r_{k,i;r}p_i(k)$ and using (\ref{systems higher recursion})
\[
 \gamma^r_1 = \sum_{k\geq 1} p^k(k) x^k - 2 \gamma^r_2 =
 \sum_{k\geq 1} p^r(k) x^k - \gamma^r_1(x)^2 - \sum_{j
      \in \mathcal{R}}s_j\gamma^j_1(x)x
    \partial_x\gamma^r_1(x)
\]
giving
\[
  \gamma^r_{1,n} = p^r(n) +
  \sum_{i=1}^{n-1}(-s_ri-1)\gamma^r_{1,i}\gamma^r_{1,n-i} +
  \sum_{\substack{j \in \mathcal{R} \\ j
      \neq r}}\sum_{i=1}^{n-1} (-s_ji)\gamma^j_{1,n-i}\gamma^r_{1,i}
\]

To attack the growth of the $\gamma^r_1$ we will again assume that
\[
   \sum_{k \geq 1} x^k \frac{p^r(k)}{k!} = f^r(x)
\]
has radius $0 < \rho_r \leq \infty$ and $f^r(x) > 0$ for $|x|\leq
\rho_r$.
We'll proceed
by similar bounds to before.

Let $a_n^r = \gamma^r_{1,n}/n!$.  Then
\[
  a_n^r = \frac{p^r(n)}{n!} +
  \sum_{i=1}^{n-1}(-s_ri-1)a^r_ia^r_{n-i}\binom{n}{i}^{-1} +
  \sum_{\substack{j \in \mathcal{R} \\ j
      \neq r}}\sum_{i=1}^{n-1} (-s_ji)a^j_{n-i}a^r_{i}\binom{n}{i}^{-1}
\]

Taking the last term in each sum we have
\[
  a_n^r \geq \frac{p^r(n)}{n!} - \left(\sum_{j \in \mathcal{R}} s_ja_1^j\right)\frac{n-2}{n}a_{n-1}^r
\]
Let $b^r_n$ be the series defined by $b^r_1 = a^r_1$ and equality in the
above recursion.  Let $\mathbf{B}^r(x) = \sum b^r_n x^n$.  Then as
before if $\sum_{j \in \mathcal{R}} s_ja_1^j = 0$ the radius of
$\mathbf{B}(x)$ is $\rho_r$ and otherwise consider
\[
  \mathbf{B}^r(x)'' = f^r(x)'' - \left(\sum_{j \in \mathcal{R}}
    s_ja_1^j\right)x\mathbf{B}^r(x)''
\]
Solving for $\mathbf{B}^r(x)''$ we get that the radius of $\sum a_n^r x^n$
is at most
\[
  \min \left\{ \rho_r, \frac{-1}{\sum_{j \in \mathcal{R}} s_ja_1^j}\right\}
\]
again interpreting the second possibility to be $\infty$ when $\sum_{j \in \mathcal{R}} s_ja_1^j=0$.

In the other direction take any $\epsilon > 0$ then there
exists an $N > 0$ such that for $n > N$ we get
\[
  a_n^r \leq \frac{p^r(n)}{n!} - \left(\sum_{j \in \mathcal{R}}
    s_ja_1^j\right)a^r_{n-1} + \epsilon \sum_{i=1}^{n-1} \sum_{j \in
    \mathcal{R}} a^r_i a^j_{n-i}
\]
Taking $\mathbf{C}^r(x)$ to be the series whose coefficients satisfy
the above recursion with equality for when this gives a result $\geq
a^r_n$ and equal to $a^r_n$ otherwise we get
\[
  \mathbf{C}^r(x) = f^r(x) - \left(\sum_{j \in \mathcal{R}}
    s_ja_1^j\right)x\mathbf{C}^r(x) + \epsilon \sum_{j \in
    \mathcal{R}}\mathbf{C}^r(x)\mathbf{C}^j(x) + P_\epsilon^r(x)
\]
where $P_\epsilon^r$ is a polynomial.

Summing over $r$ we get a recursive equation for $\sum_{r \in
  \mathcal{R}}\mathbf{C}^r(x)$ of the same form as in the single
equation case.  Note that since each $\mathbf{C}^r$ is a series with
nonnegative coefficients there can be no cancellation of singularities
and hence the radius of convergence of each $\mathbf{C}^r$ is at least
that of the sum.  The equivalent of Lemma \ref{converge} for this case
follows from
\begin{align*}
  \sum_{r \in \mathcal{R}} a_n^r & \leq \sum_{r\in\mathcal{R}}
  \frac{p^r(n)}{n!}  - \sum_{j \in \mathcal{R}}s_ja_1^j\sum_{r \in
    \mathcal{R}}a_{n-1}^r + \sum_{r, j\in\mathcal{R}}\sum_{i=1}^{n-1}(-s_j
  i)a_{n-i}^ja_i^r \binom{n}{i}^{-1} \\
  & \leq \sum_{r\in\mathcal{R}} \frac{p^r(n)}{n!}  - \sum_{j \in
    \mathcal{R}}s_ja_1^j\sum_{r \in \mathcal{R}}a_{n-1}^r \\
  & \qquad +
  \max_{j}(-s_j)\sum_{i=2}^{n-2} i \binom{n}{i}^{-1}
  \left(\sum_{r\in\mathcal{R}}a^r_{n-i}\right) \left(\sum_{r
      \in\mathcal{R}} a^r_{i}\right) \\
  & = \sum_{r\in\mathcal{R}} \frac{p^r(n)}{n!}  - \sum_{j \in
    \mathcal{R}}s_ja_1^j\sum_{r \in \mathcal{R}}a_{n-1}^r \\
  & \qquad + \max_{j}(-s_j) n
  \sum_{2 \leq i \leq \theta n} \binom{n}{i}^{-1}
  \left(\sum_{r\in\mathcal{R}}a^r_{n-i}\right) \left(\sum_{r
      \in\mathcal{R}} a^r_{i}\right) \\
  & \qquad + \max_{j}(-s_j) \frac{n}{2}
  \sum_{\theta n \leq i \leq n- \theta n} \binom{n}{i}^{-1}
  \left(\sum_{r\in\mathcal{R}}a^r_{n-i}\right) \left(\sum_{r
      \in\mathcal{R}} a^r_{i}\right)
\end{align*}
for $\theta$ as in Lemma \ref{converge} with $\sum_{r \in \mathcal{R}}
\mathbf{A}^r(x)$ in place of $\mathbf{A}(x)$, where $\mathbf{A}^r(x) =
\sum a^r(n)x^n$.  Then continue the argument as in Lemma
\ref{converge} with $\sum_{r \in
  \mathcal{R}}f^r(x)$ in place of $f(x)$ and $\max_{j}(-s_j)$ in
place of $-r$, and using the second term to get the correct linear part as
$\theta \rightarrow 0$.

Thus by the analysis of the single equation case we
get a lower bound on the radius of $\sum a^r_n x^n$ of $\min_{s \in \mathcal{R}} \{\rho_s,
-1/\sum_{j \in \mathcal{R}}s_ja_1^j\}$.  In particular if $r \in
\mathcal{R}$ is such that $\rho_r$ is minimal we see that the radius
of $\sum a^r_n x^n$ is exactly $\min\{\rho_r,
-1/\sum_{j \in \mathcal{R}}s_ja_1^j\}$.

Suppose the radius of $\sum a_n^s x^n$ was strictly greater
than that of $\sum a_n^r x^n$.  Then we can find $\beta > \delta > 0$
such that
\[
   a_n^r > \beta^n > \delta^n > a_n^s
\]
for $n$ sufficiently large.  Pick a $k \ge 1$ such that $a_k^s > 0$.  Then
\[
  \delta^n > a_n^s \geq \frac{-s_r k! a_k^s}{n\cdots (n-k+1)}
  a_{n-k}^r > \frac{-s_r k!
    a_k^s}{n\cdots (n-k+1)} \beta^{n-k}
\]
so
\[
  \frac{\delta^k}{-s_r a_k^s} \left(\frac{\delta}{\beta}\right)^{n-k}
  > \frac{k!}{n \cdots (n-k+1)}
\]
which is false for $n$ sufficiently large, giving a contradiction.
Thus all the $\sum a_n^s x^n$ have the same radius $\min_{r \in \mathcal{R}}\{\rho_r,
-1/\sum_{j \in \mathcal{R}}s_ja_1^j\}$

{}From this we can conclude
that each $\sum x^n\gamma^r_{1,n}/n!$ also converges with radius $\min_{r
  \in \mathcal{R}}\{\rho_r,
-1/\sum_{j \in \mathcal{R}}s_j\gamma_{1,1}^j\}$, where the second
possibility is interpreted as $\infty$ when $\sum_{j \in
  \mathcal{R}}s_j\gamma_{1,1}^j = 0$.

\subsection{Systems with some $s_r > 0$}
Let us relax the restriction that $s_r < 0$ and that $p^r(n) \geq 0$.  It is now difficult to
make general statements concerning the radius of convergence of the
$\sum a^r_n x^n$.  For example consider the system
\begin{align*}
  a^1_n & = \frac{p^1(n)}{n!} + \sum_{j=1}^{n-1}(2j-1)a_j^1 a_{n-j}^1
  \binom{n}{j}^{-1} - \sum_{j=1}^{n-1}ja_j^1
  a_{n-j}^2\binom{n}{j}^{-1} \\
  a^2_n & = \frac{p^2(n)}{n!} - \sum_{j=1}^{n-1}(j+1)a_j^2 a_{n-j}^2
  \binom{n}{j}^{-1} + \sum_{j=1}^{n-1}2ja_j^2
  a_{n-j}^1\binom{n}{j}^{-1}
\end{align*}
so $s_1 = -2$ and $s_2 = 1$.  Suppose also that
\begin{align*}
  p^2(2) & = 0 \\
  a_1^1 & = a^2_1 \\
  p^2(n) & = -2(n-1)!a_1^2 a_{n-1}^1 \\
\end{align*}
Then $a_2^2 = 0$ and inductively $a_n^2 = 0$ for $n\geq 2$ so the system degenerates to
\begin{align*}
  a^1_n & = \frac{p^1(n)}{n!} + \sum_{j=1}^{n-1}(2j-1)a_j^1 a_{n-j}^1
  \binom{n}{j}^{-1} - \frac{n-1}{n}a_1^1a_{n-1}^1 \\
  a^2_n & =
  \begin{cases}
    a^1_1 & \text{if n = 1} \\
    0 & \text{otherwise}
  \end{cases}
\end{align*}
We still have a free choice of $p^1(n)$, and hence control of the
radius of the $a^1$ series.  On the other hand the
$a^2$ series trivially has infinite radius of convergence.

Generally, finding a lower bound on the radii of the solution series,
remains approachable by the preceeding methods while control of the
radii from above is no longer apparent.

Precisely, for any $\epsilon > 0$
\begin{align*}
  |a_n^r|
  & \leq \frac{\left|p^r(n)\right|}{n!} + \left|-\sum_{j \in \mathcal{R}}
    s_ja_1^j\right||a^r_{n-1}| +
    \sum_{i=1}^{n-2}|(-s_ri-1)||a^r_i||a^r_{n-i}|\binom{n}{i}^{-1} \\
    & \qquad +
    \sum_{\substack{j \in \mathcal{R} \\ j
        \neq r}}\sum_{i=1}^{n-2}
    |(-s_ji)||a^j_{n-i}||a^r_{i}|\binom{n}{i}^{-1} \\
  & \leq \frac{\left|p^r(n)\right|}{n!} + \left|-\sum_{j \in \mathcal{R}}
    s_ja_1^j\right||a^r_{n-1}| + \epsilon \sum_{i=1}^{n-1} \sum_{j \in
    \mathcal{R}} |a^r_i| |a^j_{n-i}|
\end{align*}
So, for a lower bound on the radius we may proceed as in the
nonnegative case using the absolute value of the coefficients and
achieving that the radius of the $\sum a_n^r x^n$,  and hence that of
$\sum x^n\gamma^r_{1,n}/n!$, is at least
\[
  \min_{r \in \mathcal{R}}\left\{\rho_r,
  \left|\frac{-1}{\sum_{j \in \mathcal{R}}s_j\gamma_{1,1}^j}\right|\right\}
\]
where the second possibility is interpreted as $\infty$ when
$\sum_{j \in \mathcal{R}}s_j\gamma_{1,1}^j = 0$.

Note that this gives the lower bound on the radius of convergence as
the minimum of the first instanton singularity (which one expects to
be the radius for $p(k)/k!$) and the inverse of the first term in
the $\beta$ function of the theory. Furthermore, we emphasize that
Ward identities typically allow a restriction to systems where all
$s_r<0$. A more detailed discussion will be given in future work
where the general approach described here will be discussed with
regard to the specific details of the relevant renormalizable
theories of interest. Finally, we note that the apperance of the
inverse of the first term in the $\beta$-function makes sense: in
the conformal case of a vanishing $\beta$-function we would not
expect a constraint on the minimum of the radius to come from
perturbation theory.
\section{Applications of the growth of $\gamma_1$}
While expectations for the growth of $p(k)/k!$ in terms of instanton singularities are routine in the context of path integral estimates,
the path integral is merely a successful heuristic to parameterize our lack of understanding of quantum field theory. Rigorous estimates
for the growth of superficially convergent Green functions, and hence the $p(k)$, can sometimes be obtained using constructive field theory,
at least as bounds for the radius \cite{Riv}. We emphasize that such results can be turned by our methods into similarly rigorous results for superficially divergent
Green functions.

A more complete discussion, dedicated to the renormalizable quantum field theories in four dimensions, will be given elsewhere.
\appendix
\section{Other theories combinatorially with one primitive per loop
  order}\label{others}
For each of the following systems the solution series have
coefficients satisfying the universal law.  In the mixed $\phi^3$,
$\phi^4$ case there is one primitive per vertex per loop order.
Unfortunately the full power of symmetry factors is not available in
this simple combinatorial set-up, leading to the different variants.

\begin{tabular}{lll}
Theory & System & Radius \\
\hline
$\phi^3$ & $\begin{aligned}
  X_1 & = 1+X_1F_1 \\
  F_1 & = xX_1^2F_2^3(1+F_1) \\
  X_2 & = \frac{xX_1F_2^2}{2} \\
  F_2 & = 1 + F_2X_2\end{aligned}$& \parbox{4.5cm}{Smallest positive
  real root of $3581577 x^4  - 4443984 x^3  + 2332368 x^2  - 539136 x
  + 32768$.  Numerically $0.09061681898407704\ldots$}\\
\hline
$\phi^4$ & $\begin{aligned}
  X_1 & = 1 + x^2X_1^3F_2^4F_1 + \frac{3}{2}xX_1F_2^2 \\
  F_1 & = xX_1F_2^2(1+F_1) \\
  X_2 & = xF_2 + x^2X_1F_2^3 \\
  F_2 & = 1 + F_2X_2
\end{aligned}$
& \parbox{4.5cm}{Root of a degree 10 polynomial. Numerically $0.12968592295019730\ldots$}\\
\hline
$\phi^4$ variant & $\begin{aligned}X_1 & = 1 + x^2X_1^3F_2^4F_1 +
  xX_1F_2^2 \\ &\text{rest as in previous case}\end{aligned}$ &
\parbox{4.5cm}{Root of a degree 9 polynomial. Numerically
  $0.13856076790723086\ldots$}\\
\hline
mixed $\phi^3$, $\phi^4$ & $\begin{aligned}
  X_1  = & 1 + xX_1X_2F_3^2 + xX_1^2F_3^3 \\
  & + 2xF_3^2X_1 \\
  & + X_2X_1^2F_3(F_aF_b + F_a + F_b \\
  & \qquad + F_axX_2^2F_3^3 + F_bxX_2F_3^2 \\
  & \qquad + x^2X_1^2X_2F_3^5 + xX_1^2F_3^3 \\
  & \qquad + xX_2F_3^2) \\
  X_2 = & 1 + X_2(F_aF_b + F_a + F_b \\
  & \qquad + F_a xX_1^2F_3^3 + \\
  & \qquad F_b xX_2F_3^2 + x^2X_1^2X_2F_3^5) \\
  & + X_1^3F_3(F_aF_b +F_a + F_b \\
  & \qquad + F_a xX_1^2F_3^3 + F_b xX_2F_3^2 \\
  & \qquad + x^2X_1^2X_2 F_3^5 \\
  & \qquad + xX_1^2F_3^3 + xX_2F_3^2) \\
  &+ \frac{3}{2}xX_2F_3^2 \\
  F_a = & x^2X_2^2F_3^4 + xX_2F_3^2F_a \\
  F_b = & x^2X_1^4F_3^6 + xX_1^2F_3^3F_b \\
  X_3 = & xX_1F_3^2 + xF_3 + x^2X_2F_3^3 \\
  F_3 = & 1 + X_3F_3
\end{aligned}
$& 0.02145\ldots
\end{tabular}


\begin{thebibliography}{99}
\bibitem{univlaw}Jason Bell, Stanley Burris, and Karen Yeats,
  \emph{Counting Rooted Trees: The Universal Law $t(n) \sim C
    \rho^{-n} n^{-3/2}$}.
  Elec. J. Combin.
  \textbf{13}
  [arxiv:math.CO/0512432].
\bibitem{BKW} Isabella Bierenbaum, Dirk Kreimer, Stefan Weinzierl, \emph{The next to leading ladder approximation}, hep-th/0612180.

\bibitem{can}E.~Rodney Canfield, \emph{Remarks on an asymptotic
  method in combinatorics}. J.~Combin.~Theory Ser.~A \textbf{37}
  (1984), no. 3, 348--352.
\bibitem{cvitanovic}Predrag Cvitanovi\'c, \emph{Asymptotic estimates
    and gauge invariance}.  Nucl.~phys.,~B \textbf{127} (1977) 176--188.
\bibitem{drm}Micheal Drmota, \emph{Combinatorics and asymptotics on
  trees}. Cubo Journal, \textbf{6}, no. 2, (2004)
\bibitem{drsys}Micheal Drmota, \emph{Systems of functional
  equations}. Random Struct.~Alg., \textbf{10}, (1997), 103--124
\bibitem{fs}Philippe Flajolet and Robert Sedgewick,
  \emph{Analytic Combinatorics.} Online draft available at
  http://algo.inria.fr/flajolet/Publications/books.html.
\bibitem{hrs}F.~Harary, Robert W.~Robinson and Allen J.~Schwenk,
  \emph{Twenty-step algorithm for determining the asymptotic
    number of trees on various species.}
  J.~Austral.~Math.~Soc. (Ser.~A) \textbf{20} no. 4 (1975),
  483--503. \emph{Corrigendum:} J.~Austral.~Math.~Soc. (Ser.~A)
  \textbf{41} no. 3 (1986), 325.
\bibitem{jbw} K.~Johnson, M.~Baker, and R.~Willey, \emph{Self-Energy
    of the Electron}. Phys.~Rev. \textbf{136}, 4B, (1964), B1111--B1119.
    \bibitem{OnOverl} %\cite{Kreimer:1998iv}
  Dirk Kreimer,
  \emph{On overlapping divergences,}
  Commun.\ Math.\ Phys.\  {\bf 204}, 669 (1999)
  [arXiv:hep-th/9810022].
  %%CITATION = HEP-TH 9810022;%%
  \bibitem{dirkLL}
%\cite{Kreimer:2002fv}
  Dirk Kreimer,
  \emph{Unique factorization in perturbative QFT,}
  Nucl.\ Phys.\ Proc.\ Suppl.\  {\bf 116}, 392 (2003)
  [arXiv:hep-ph/0211188];\\
  %%CITATION = HEP-PH 0211188;%%
\emph{\'Etude for linear Dyson Schwinger Equations}, IHES preprint 2006, http://www.ihes.fr/PREPRINTS/2006/P/P-06-23.pdf.
\bibitem{old}
%\cite{Kreimer:2002rf}
  Dirk Kreimer,
  \emph{Structures in Feynman graphs: Hopf algebras and symmetries,}
  Proc.\ Symp.\ Pure Math.\  {\bf 73}, 43 (2005)
  [arXiv:hep-th/0202110].
  %%CITATION = HEP-TH 0202110;%%
\bibitem{anatomy} Dirk Kreimer, \emph{Anatomy of a gauge theory},
   Annals Phys.\  {\bf 321}, 2757 (2006)
  [arXiv:hep-th/0509135].
  %%CITATION = HEP-TH 0509135;%%

\bibitem{dktor} %\cite{Kreimer:2006va}
  Dirk Kreimer,
  \emph{Dyson Schwinger equations: From Hopf algebras to number theory,}
  arXiv:hep-th/0609004.
  %%CITATION = HEP-TH 0609004;%%
\bibitem{non lin} Dirk Kreimer and Karen Yeats, \emph{An \'Etude in
    non-linear Dyson-Schwinger Equations,} Nucl.~Phys.~B Proc.~Suppl., \textbf{160}, (2006), 116--121. [arxiv:hep-th/0605096].
\bibitem{lal}Steven P.~Lalley, \emph{Finite range random walk on free
  groups and homogeneous trees}, Ann.~Probab., \textbf{21}, no. 4
  (1993), 2087--2130
\bibitem{mm}A.~Meir and J.~W.~Moon, \emph{On an asymptotic method
  in enumeration.} J.~Combin.~Theory Ser.~A  \textbf{51} (1989),
  no. 1, 77--89.  \emph{Erratum:} J.~Combin.~Theory Ser.~A
  \textbf{52} (1989), no. 1, 163.
\bibitem{odly}A.~M.~Odlyzko, {\em Asymptotic enumeration methods.}
  Handbook of combinatorics, \textbf{Vol. 1, 2}, 1063--1229, Elsevier,
  Amsterdam, 1995.
\bibitem{otter}R.~Otter, \emph{The number of trees.}
  Annals of Mathematics \textbf{49} (1948), 583--599.
\bibitem{po}G.~P\'olya, \emph{Kombinatorische Anzahlbestimmungen f\"ur
  Gruppen, Graphen und chemische Verbindungen}. Acta Math. \textbf{68}
  (1937), 145--254.
\bibitem{pr}G.~P\'olya and R.~C.~Read, \emph{Combinatorial
  enumeration of groups, graphs, and chemical
  compounds}. Springer-Verlag, New York, 1987.
\bibitem{OEIS}N.~J.~A.~Sloane, The On-Line Encyclopedia of
  Integer Sequences, published on-line at
  www.research.att.com/$\sim$njas/sequences/
  (2006).
\bibitem{WvS1} %\cite{vanSuijlekom:2006ig}
  Walter D.~van Suijlekom,
  \emph{The hopf algebra of Feynman graphs in QED,}
  Lett.\ Math.\ Phys.\  {\bf 77}, 265 (2006)
  [arXiv:hep-th/0602126].
  %%CITATION = HEP-TH 0602126;%%
\bibitem{WvS2}
%\cite{vanSuijlekom:2006fk}
  Walter D.~van Suijlekom,
  \emph{Renormalization of gauge fields: A Hopf algebra approach,}
  arXiv:hep-th/0610137.
  %%CITATION = HEP-TH 0610137;%%

\bibitem{wo}Alan R.~Woods, \emph{Coloring rules for finite trees,
  probability of monadic second order sentences}, Random Struct.
  Alg., \textbf{10}, (1997), 453--485.
  \bibitem{Riv}%\cite{Magnen:1987gy}
%\bibitem{Magnen:1987gy}
  J.~Magnen, F.~Nicolo, V.~Rivasseau and R.~Seneor,
  \emph{A Lipatov bound for $\phi^4$ in four-dimensions Euclidean field theory,}
  Commun.\ Math.\ Phys.\  {\bf 108} (1987) 257.
  %%CITATION = CMPHA,108,257;%%
\end{thebibliography}
\end{document}